\documentclass[aps,prl,twocolumn,reprint,superscriptaddress]{revtex4-2}
\usepackage{graphicx}
\usepackage{amsmath}
\usepackage{amssymb}
\usepackage{colordvi}
\usepackage{mathrsfs}
\usepackage{bm}
\usepackage{verbatim}
\usepackage{dcolumn}
\usepackage{epsfig}
\usepackage{subfigure}
\usepackage[colorlinks,allcolors=blue]{hyperref}
\usepackage{ulem}
\usepackage{dsfont}
\usepackage{makecell}
\usepackage{tikz}
\usepackage{lipsum,epsfig,dsfont}
\begin{document}
\title{Fusion Rules of Majorana-Kramer-Pairs in Time-Reversal-Invariant Topological Superconductors}
\author{Hongfa Pan}
\affiliation{International Centre for Quantum Design of Functional Materials, CAS Key Laboratory of Strongly-Coupled Quantum Matter Physics, and Department of Physics, University of Science and Technology of China, Hefei, Anhui 230026, China}
\author{Jinxiong Jia}
\affiliation{International Centre for Quantum Design of Functional Materials, CAS Key Laboratory of Strongly-Coupled Quantum Matter Physics, and Department of Physics, University of Science and Technology of China, Hefei, Anhui 230026, China}	
\author{Zhenhua Qiao}
\email[Correspondence author:~]{qiao@ustc.edu.cn}
\affiliation{International Centre for Quantum Design of Functional Materials, CAS Key Laboratory of Strongly-Coupled Quantum Matter Physics, and Department of Physics, University of Science and Technology of China, Hefei, Anhui 230026, China}
\affiliation{Hefei National Laboratory, University of Science and Technology of China, Hefei 230088, China}
\date{\today{}}
\begin{abstract}
 We theoretically investigate the fusion rules of Majorana Kramers pairs in time-reversal-invariant topological superconductors. We find that the fusion of Majorana Kramers pairs is a process that Ising anyons fuse independently in the two distinct time-reversal sectors. Considering the full fusion including the initialization and the fusion, we explore the observation of a supersymmetry that emerges in time-reversal-invariant topological superconductors, and design the schemes for the nontrivial fusion and the trivial fusion to show the non-Abelian statistics of Majorana Kramers pairs. We also show the possible influence of local adiabatic mixing on the fusion and the differentiation between distinct fusion processes remains feasible even in the presence of such mixing. Our proposals are applied in $d_{x^2-y^2}$-wave topological superconductors, and the theoretical framework can be extended to the fusion of multiple Majorana zero modes protected by unitary symmetry.
\end{abstract}

\maketitle
\textit{Introduction---.} The profound implications for fault-tolerant topological quantum computing have engendered extensive research interest in topological superconductors~\cite{C. Nayak2008,Alicea2011,Alicea2012,Sato2017,M.Sato2016, C.W2013,X.-J. Liu2020,Kitaev2001,M. Franz2015,Green2000}. As quasi-particles, Majorana zero modes (MZMs)~\cite{L. Fu2008,R. M. Lutchyn2010,M. Sato2009,F. Pientka2017,Kitaev2001,Y. Oreg2010,P. W. Brouwer2017,F. Zhang2018} and Majorana Kramers pairs (MKPs)~\cite{C. L. M. Wong2012,Qi2009,F. Zhang2013,Jelena Klinovaja2020,Karsten Flensberg2014,D. Loss2018,E. Berg2013} exhibit the statistical properties of non-Abelian anyons at the boundaries of topological superconductors (TSCs) in the absence and presence of time-reversal symmetries $\mathcal{T}$, respectively. Braiding~\cite{Alicea2011,X.-J. Liu2014,J. Liu2021,Corneliu Malciu2018,Chetan Nayak2008,Liang Fu2022,Xiaoyu Zhu2018,Y. Tanaka2022} and nontrivial fusion~\cite{Alicea2011,D. Aasen2016,Tong Zhou2022,E. Altman2015,C. Nayak2008} exhibit two distinct manifestations of non-Abelian anyons, which involve the exchange of MZMs (MKPs) and the multiple
fusion channels~\cite{C. Nayak2008}, respectively. It is well-known that definitive experimental confirmation of TSCs should involve both braiding and nontrivial fusion of MZMs or MKPs. So far, most experimental signals~\cite{K. T. Law2009,H.Shtrikman2012,J. Liu2012,L. P. Kouwenhoven2012}, such as zero-bias conductance peaks, cannot provide convincing evidence of MZMs or MKPs. 
	
In time-reversal broken TSCs, MZM is analogous to an Ising anyon, and a variety of theoretical frameworks and experimental proposals have been proposed for braiding and fusion~\cite{Alicea2011,J. Liu2021,D. Aasen2016,Tong Zhou2022}. However, the fusion rules of Majorana Kramer pairs remain unclear in time-reversal-invariant TSCs. Since the suppression of superconducting energy gap can be avoided due to the absence of applying magnetic field~\cite{C. L. M. Wong2012,Qi2009,F. Zhang2013}, thus it possesses great potential in the possible experimental realization. Furthermore, it is clear that fusion is much easier than braiding for MKPs and some schemes of braiding are even based on fusion~\cite{F. von Oppen2107,Liang Fu2022,Chetan Nayak2008}. Therefore, it is crucial and meaningful to study the fusion rules for MKPs.

In this Letter, we show that the fusion of two MKPs can be divided into two progresses in the different time-reversal sectors, corresponding to their braiding~\cite{X.-J. Liu2014,Y. Tanaka2022}. For the fusion of two MKPs related by $\mathcal{T}$, we design a scheme to observe the emergence of supersymmetry $\mathcal{T}$~\cite{Qi2009} in one-dimensional time-reversal-invariant TSCs. Based on the independent fusion of two time-reversal sectors, one can construct an effective model formed by four MKPs with tunable couplings to investigate the fusion rules. We further design a scheme to realize the nontrivial and trivial fusions, showing multiple fusion channels and behaving as a controlled experiment, respectively. In particular, MKPs can fuse to four states including odd and even fermion-parity with an equal probability of $25\%$ in the nontrivial fusion, while MKPs fuse to the ground state with an absolute probability of $100\%$ in the trivial case.

Local adiabatic mixing can occur in time-reversal-invariant TSCs, which may bring controversial influence on the braiding of MKPs~\cite{Wolms2014,Wolms2016,P. Gao2016}. However, in most cases, we demonstrate that such mixing can be negligibly small within the fusion dynamics as a higher-order effect. Even though this mixing is included, we find that the trivial and nontrivial fusions can still be well distinguished as the mixing is in the same fermion-parity states. Therefore, this provides a special method to confirm the realization of time-reversal-invariant TSCs by introducing the mixing artificially. In the end, we show that both the trivial and nontrivial fusions can be realized in the concrete one-dimensional $d_{x^2-y^2}$-wave TSCs~\cite{Y. Tanaka2022,C. L. M. Wong2012}. The results can be measured by the fractional Josephson effect~\cite{,X.-J. Liu2014,F.Zhang2014,F. Zhang2013,B. Béri,F. von Oppen2017} and so on. Encouragingly, our theoretical framework can be naturally extended to the fusion of MZMs protected by unitary symmetry~\cite{Xiong-Jun Liu2022}.

\begin{figure}
	\centering 
	\includegraphics[width=0.49\textwidth]{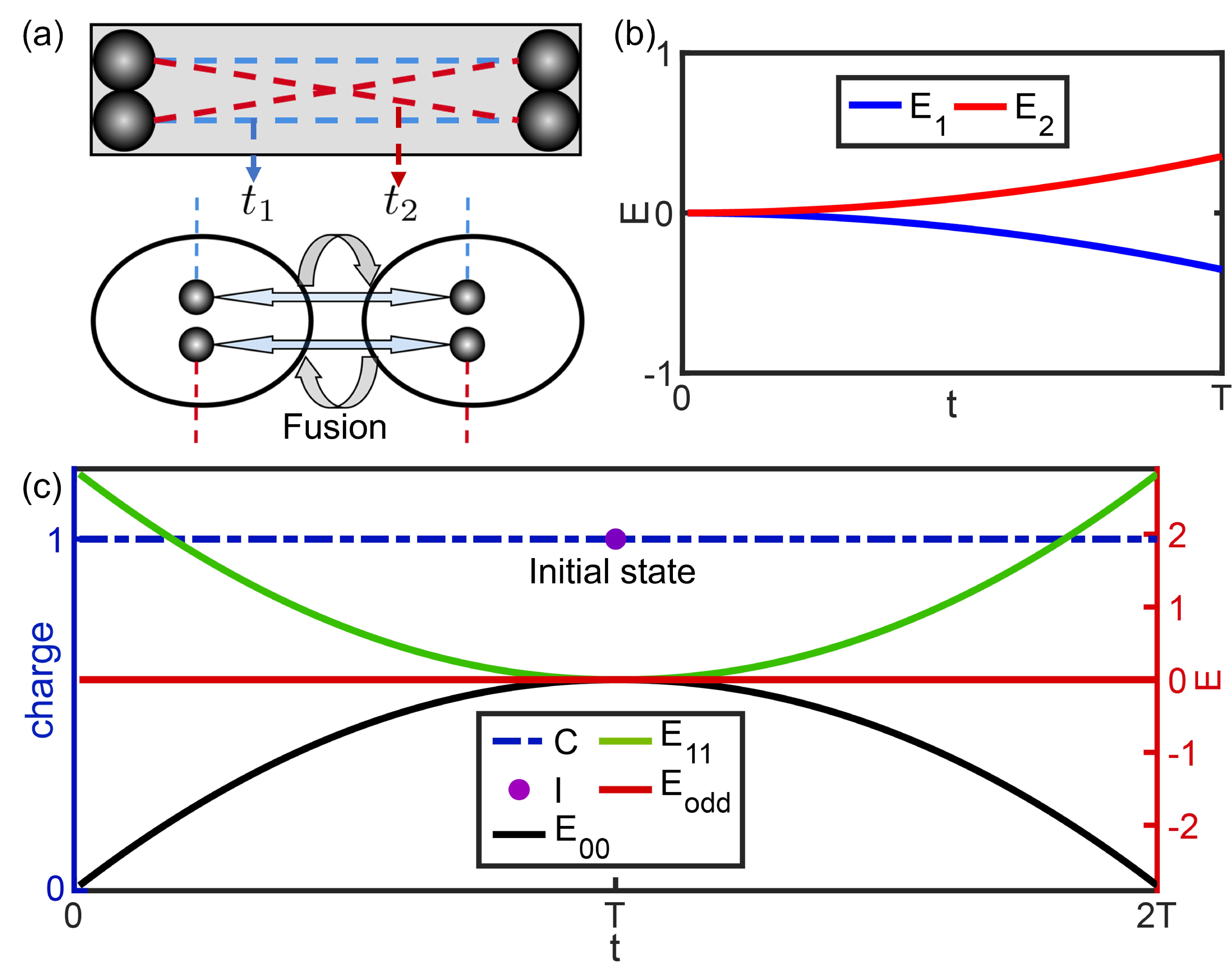} 
	\caption{(a) Two MKPs at two ends of a 1D TSC with $\mathcal{T}$ (upper) and the fusion of two MKPs (lower). (b) Evolution of energy eigenvalues as a function of the couplings with T=100. (c) Evolution of energy eigenvalues and charge as a function of the couplings. The blue line C denotes charge evolution. The lines $E_{11}$ and $E_{00}$ denotes the energy evolution of $|0\tilde{0}\rangle$ and $|1\tilde{1}\rangle$, while the line $E_{odd}$ denotes the energy evolution of
		$|0\tilde{1}\rangle$ or $|1\tilde{0}\rangle$. In our calculation, the couplings evolve as a quadratic function. $t_{1,2}$ are initially set to be $-1$, $T_{c}=-1$ and $T=1000$.}
	\label{fig1}.
\end{figure}

\textit{Independent fusion and supersymmetry---.} The MKPs formed by $\gamma_{L}$ ($\gamma_{R}$) and $\tilde{\gamma}_{L}$ ($\tilde{\gamma}_{R}$) at the left (right) end in a 1D TSC with $\mathcal{T}^{2}=-1$, with the symmetry allowed couplings $t_{1}$ and $t_{2}$ that are denoted by blue and red dashed lines, respectively, are shown in the upper panel of Fig.~\ref{fig1}(a). Fusing them to bound states can be described by the effective Hamiltonian: 
\begin{equation} 
H_{0}=it_{1}(\gamma_{L}\gamma_{R}+\tilde{\gamma}_{L}\tilde{\gamma}_{R})+it_{2}(\gamma_{L}\tilde{\gamma}_{R}-\tilde{\gamma}_{L}\gamma_{R}). 
\end{equation} 
Supposing that $t_{1}$ and $t_{2}$ are evolved from $-0.5$ to $0$ as a quadratic functional form, the evolution of two energy values with twofold degeneracy related by $\mathcal{T}$ is shown in Fig.~\ref{fig1}(b). Analytically, two new operators $\gamma_{1}$ and $\gamma_{2}$ can be defined in the fusion, as well as their time reversal partners $\tilde{\gamma}_{1}$ and $\tilde{\gamma}_{2}$, so that $H_{0}=iE_{0} (\gamma_{1L}\gamma_{1R}+\tilde{\gamma}_{1L}\tilde{\gamma}_{1R})$~\cite{SM}. On one hand, two MKPs can fuse into the electronic bound states denoted by two electron creation operations, which are defined as $c^{\dagger} = \frac{1}{2}(\gamma_{1L} + i\gamma_{1R})$ and $\tilde{c}^{\dagger} = \frac{1}{2}(\tilde{\gamma}_{1L} + i\tilde{\gamma}_{1R})$, respectively. This indicates that the fusion is independent of the two different time-reversal sectors in an adiabatic evolution. Thus, the Majorana end modes can fuse independently as indicated by the world lines with different colors as plotted in the lower panel of Fig.~\ref{fig1}(a).
On the other hand, two decoupled MKPs with $i\gamma_{L}\tilde{\gamma}_{L}=-1$ and $i\gamma_{R}\tilde{\gamma}_{R}=1$, respectively, are related by $\mathcal{T}$. As changing the fermion number by an odd number, $\mathcal{T}$ emerges as a supersymmetry~\cite{Qi2009}. It offers the opportunity to experimentally observe the supersymmetry in TSCs with $\mathcal{T}$. We can then define another two electron creation operations $f^{\dagger}=1/2(\gamma_{L}+i\gamma_{R})$ and $\tilde{f}^{\dagger}=1/2(\tilde{\gamma}_{L}+i\tilde{\gamma}_{R})$. The resulting Hamiltonian becomes $H^{\prime}_{0}=2t_{1}(1-f^{\dagger}f-\tilde{f}^{\dagger}\tilde{f})+2t_{2}(f^{\dagger}\tilde{f}^{\dagger}-f\tilde{f})$. On the basis of $\{|0_{f}\tilde{0}_{f}\rangle, f^{\dagger}\tilde{f}^{\dagger}|0_{f}\tilde{0}_{f}\rangle, f^{\dagger}|0_{f}\tilde{0}_{f}\rangle, \tilde{f}^{\dagger}|0_{f}\tilde{0}_{f}\rangle\}$, the Hamiltonian $H^{\prime}_{0}$ can be expressed in a matrix form~\cite{SM}.  

To observe the emergent supersymmetry $\mathcal{T}$, there are two steps including the initialization and the fusion. At the initialization, we choose the odd fermion-parity state $|0\tilde{1}\rangle$ or $|1\tilde{0}\rangle$, which are eigenvectors of $H_{0}$. They are degenerate as shown in the line $E_{odd}$, indicating the evolution of the energy eigenvalues of the odd fermion-parity states in Fig.~\ref{fig1}(c). The initialization can be set by evolving $t_{1}$ and $t_{2}$ to $0$ adiabatically, as shown at the purple dot I in the blue line, which indicates the evolution of the charge. The evolution of the energy eigenvalues of $|0\tilde{0}\rangle$ and $|1\tilde{1}\rangle$ are displayed in the black line $E_{00}$ and the green line $E_{11}$, respectively. At $t=T$, two decoupled MKPs emerge with $i\gamma_{L}\tilde{\gamma}_{L}=-1$ ($i\gamma_{L}\tilde{\gamma}_{L}=1$) and $i\gamma_{R}\tilde{\gamma}_{R}=1$ ($i\gamma_{R}\tilde{\gamma}_{R}=-1$)~\cite{SM}, respectively. The second step, i.e., fusion, can then be manipulated by evolving $t_{1}$ and $t_{2}$ to $T_{c}$ adiabatically. A fermion with one charge is left finally, which can be regarded as the observation of supersymmetry.
\begin{figure*}
	\centering  
	\includegraphics[width=0.98\textwidth]{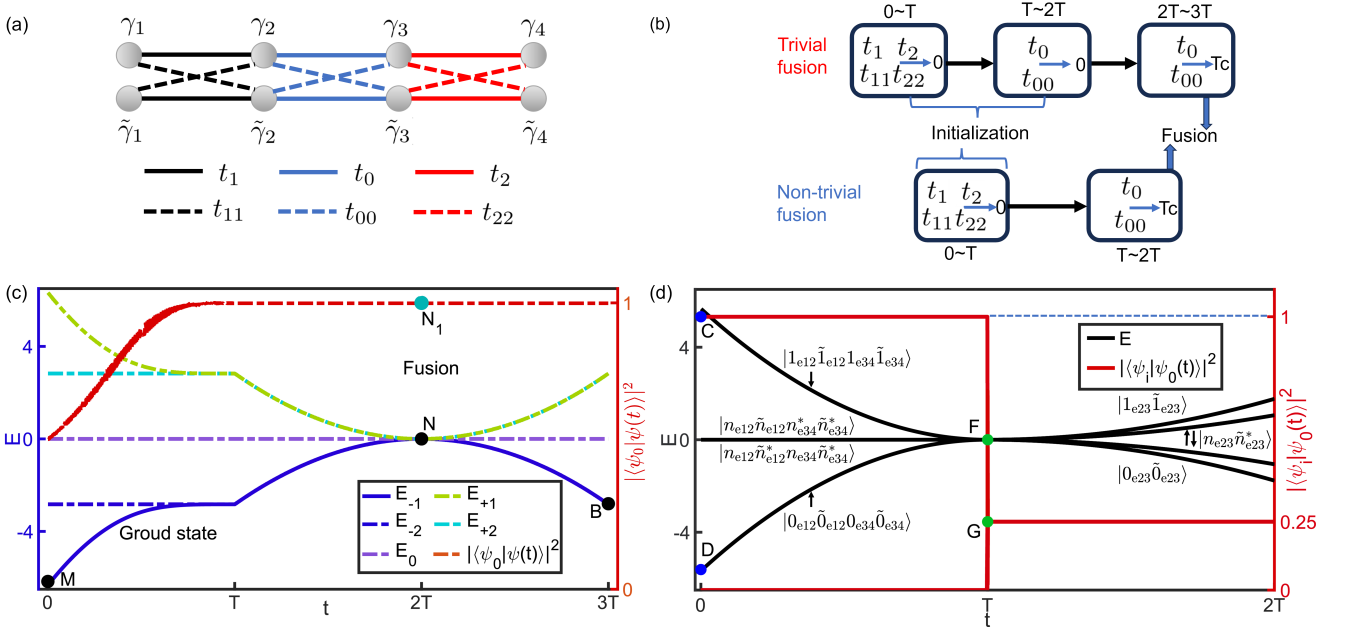} 
	\caption{(a) The effective model's construction. (b) The manipulation steps in the trivial and nontrivial fusions. Assuming the couplings evolve adiabatically in a quadratic function form. (c) The evolution of wave function and eigenvalues in the trivial fusion. The ground state initially denoted by the dot M and the state returns to the ground state, as indicated by the dot B. In our calculation, $T_{c}=-1$ and $T=2000$. (d) The evolution of wave function and energy eigenvalues in the nontrivial fusion. The system is initially in the ground state denoted by the dot D and the dot C. By fusion, the state evolves into four states $|0_{\mathrm{e23}}\tilde{0}_{\mathrm{e23}}\rangle$, $|1_{\mathrm{e23}}\tilde{1}_{\mathrm{e23}}\rangle$, $|1_{\mathrm{e23}}\tilde{0}_{\mathrm{e23}}\rangle$ and $|0_{\mathrm{e23}}\tilde{1}_{\mathrm{e23}}\rangle$ with an equal probability $25\%$. In our calculation,  $T=1000$ and $T_{c}=-1$ for the couplings $i\gamma_{2}\gamma_{3}$ and $i\gamma_{2}\tilde{\gamma}_{3}$, while $T_{c}=-0.5$ for the couplings $i\tilde{\gamma}_{2}\tilde{\gamma}_{3}$ and $i\tilde{\gamma}_{2}\gamma_{3}$.}       
	\label{fig2}.
\end{figure*} 

\textit{Nontrivial and trivial fusions---.} Let us now move to discuss the nontrivial fusion in a time-reversal invariant TSC with $\mathcal{T}^{2}=-1$. MZMs behave as Ising anyons in TSCs without $\mathcal{T}$, obeying the fusion rule of $\gamma\times\gamma=I+\psi$, which indicates that two MZMs can fuse into a trivial particle $I$ (i.e., they can annihilate to the vacuum) or a fermion $\psi$~\cite{C. Nayak2008,D. Aasen2016}. As the fusion is independent in two different time-reversal sectors, MKPs behave as Ising anyons in each sector, obeying the fusion rules of $\gamma\times\gamma=I+\psi$ and $\tilde{\gamma}\times\tilde{\gamma}=I+\tilde{\psi}$, respectively. In Ising anyons theory, the symbol of $F^{\sigma\sigma\sigma}_{\sigma}$~\cite{C. Nayak2008,D. Aasen2016} describes the transition of the basis vectors among four Ising anyons. Due to $\mathcal{T}$, $\tilde{F}^{\tilde{\sigma}\tilde{\sigma}\tilde{\sigma}}_{\tilde{\sigma}}$ is necessary to describe the transition in its time-reversal sector. Specifically, we use effective model to study the corresponding fusion rules as displayed in Fig.~\ref{fig2}(a). The different couplings satisfying $\mathcal{T}$ are respectively denoted by the solid and dashed lines. The Hamiltonian is written as: 
\begin{equation}
	\begin{aligned}
		H_{e}&=it_{1}(\gamma_{1}\gamma_{2}+\tilde{\gamma}_{1}\tilde{\gamma}_{2})+it_{2}(\gamma_{1}\tilde\gamma_{2}-\tilde{\gamma}_{1}\gamma_{2})\\
		&+it_{11}(\gamma_{3}\gamma_{4}+\tilde{\gamma}_{3}\tilde{\gamma}_{4})+it_{22}(\gamma_{3}\tilde\gamma_{4}-\tilde{\gamma}_{3}\gamma_{4})\\
		&+it_{0}(\gamma_{2}\gamma_{3}+\tilde{\gamma}_{2}\tilde{\gamma}_{3})+it_{00}(\gamma_{2}\tilde\gamma_{3}-\tilde{\gamma}_{2}\gamma_{3}).
	\end{aligned}
\end{equation}
Working in the even-parity subspace, the Hamiltonian $H_{e}$ can be rewritten in a matrix form~\cite{SM}. Two steps, i.e., the initialization and the fusion, are needed to realize both trivial and nontrivial fusions. The couplings, i.e., $t_{1,11}$,$t_{2,22}$ and $t_{0,00}$, are set to be $-1$ initially and \textit{evolved} adiabatically following a quadratic function as displayed in Fig.~\ref{fig2}(b), except that $t_{0,00}$ is set to be 0 for the nontrivial fusion. Thus, by these steps, one can bind $\gamma_{1,4}$ ($\gamma_{1,2}$) and $\tilde{\gamma}_{1,4}$ ($\tilde{\gamma}_{1,2}$) out of the vacuum from $0$ to $2T$ ($0$ to $T$), similarly for $\gamma_{2,3}$ ($\gamma_{3,4}$) and $\tilde{\gamma}_{2,3}$ ($\tilde{\gamma}_{3,4}$) in the trivial fusion (nontrivial fusion), then fuse $\gamma_{2,3}$ and $\tilde{\gamma}_{2,3}$ finally.

In the trivial fusion, there are five distinct energy eigenvalues to denote the states $|n_{\mathrm{e12}}\tilde{n}_{\mathrm{e12}}n_{\mathrm{e34}}\tilde{n}_{\mathrm{e34}}\rangle$, i.e., $n_{\mathrm{e12}}=0, 1$ and $\tilde{n}_{\mathrm{e12}}=0, 1$, which are the eigenvectors of the Hamiltonian $H_{e}$, as displayed in Fig.~\ref{fig2}(c) by the lines $E_{\pm 1}$, $E_{\pm 2}$, and $E_{0}$. At zero temperature, the system is theoretically at the ground state of $|0_{\mathrm{e12}}\tilde{0}_{\mathrm{e12}}0_{\mathrm{e34}}\tilde{0}_{\mathrm{e34}}\rangle$, which is denoted by the dot M. As the couplings $t_{1, 11}$ and $t_{2, 22}$ are evolved to 0 firstly, then $t_{0, 00}$ is evolved to 0, the system becomes initialized as denoted by dot N and $\rm{N_{1}}$ at the time of $t=2T$. As the couplings $t_{0}$ and $t_{00}$ are evolved to $T_{c}$, decoupled $\gamma_{2,3}$ and $\tilde{\gamma}_{2,3}$ reach the trivial fusion. In the full progress, the evolution of the state is shown by the red line $|\langle\psi_{0}|\psi(t)\rangle|^{2}$, where $\psi(t)$ and $\psi_{0}$ are the initial ground states that evolve as $H_{e}$ and the instantaneous ground state at $t=3T$, respectively. This state subsequently fuse back to the ground state $\psi_{0}$, represented by dot B, as indicated by $|\langle\psi_{0}|\psi(t)\rangle|^{2}=1$ at the end of the fusion as displayed in panel (c).

In the nontrivial fusion, the initial setting of the couplings results that $|0_{\mathrm{e12}}\tilde{0}_{\mathrm{e12}}0_{\mathrm{e34}}\tilde{0}_{\mathrm{e34}}\rangle$ and $|1_{\mathrm{e12}}\tilde{1}_{\mathrm{e12}}1_{\mathrm{e34}}\tilde{1}_{\mathrm{34}}\rangle$ are separated by a gap, while the states $|n_{\mathrm{e12}}\tilde{n}_{\mathrm{e12}}n^{*}_{\mathrm{e34}}\tilde{n}^{*}_{\mathrm{e34}}\rangle$ and $|n_{\mathrm{e12}}\tilde{n}^{*}_{\mathrm{e12}}n_{\mathrm{e34}}\tilde{n}^{*}_{\mathrm{e34}}\rangle$ (i.e., $n$,$n^{*}$=0,1 or 1,0) exhibit a sextuple degeneracy, shown by the lines $E$ in Fig.~\ref{fig2}(d). Theoretically, the system is in the ground state $|0_{\mathrm{e12}}\tilde{0}_{\mathrm{e12}}0_{\mathrm{e34}}\tilde{0}_{\mathrm{e34}}\rangle$, as indicated by the blue dot D at zero temperature. At the initialization, the couplings $t_{1,11}$ and $t_{2,22}$ are evolved to $0$ adiabatically. As the evolution begins at dot C, the state remains in the ground state as shown by the part that $|\langle\psi_{i}|\psi_{0}(t)\rangle|^{2}=1$ from $t=0$ to $t=T$, where $|\psi_{i}\rangle$ denote all the different instantaneous eigenstates at $t=0$ and $|\psi_{0}(t)\rangle$ is the ground state that evolves as $H_{e}$. At $t=T$, the initial state needed in the nontrivial fusion is well prepared, denoted by dots F and G that the zero energy states correspond to the decoupled MKPs as displayed in panel (d). After this, one can fuse the decoupled $\gamma_{2,3}$ and $\tilde{\gamma}_{2,3}$. Corresponding to the experimental measurement, we destroy $\mathcal{T}$ by taking different values of $t_{0}$ in the couplings $i\gamma_{2}\gamma_{3}$ and $i\tilde{\gamma}_{2}\tilde{\gamma}_{3}$, as well as $t_{00}$ in the couplings $i\gamma_{2}\tilde{\gamma}_{3}$ and $i\tilde{\gamma}_{2}\gamma_{3}$. After fusing by evolving $t_{0}$ and $t_{00}$ to different values of $T_{c}$ in these terms, the state evolves into a superposition state, shown by the evolution of the wave function $|\langle\psi_{i}|\psi_{0}(t)\rangle|^{2}$, where $|\psi_{i}\rangle$ denotes all the instantaneous eigenstates at $t=2T$. Because of the local measurement and the decoupling of $\gamma_{1, 4}$ and $\tilde{\gamma}_{1, 4}$ during the fusion, the eigenstates can be expressed as $|0_{\mathrm{e23}}\tilde{0}_{\mathrm{e23}}\rangle$, $|1_{\mathrm{e23}}\tilde{1}_{\mathrm{e23}}\rangle$, and $|n_{\mathrm{e23}}\tilde{n}^{*}_{\mathrm{e23}}\rangle$. As shown in panel (d), $\gamma_{2,3}$ and $\tilde{\gamma}_{2,3}$ fuse into $|0_{\mathrm{e23}}\tilde{0}_{\mathrm{e23}}\rangle$, $|1_{\mathrm{e23}}\tilde{1}_{\mathrm{e23}}\rangle$, $|1_{\mathrm{e23}}\tilde{0}_{\mathrm{e23}}\rangle$ and $|0_{\mathrm{e23}}\tilde{1}_{\mathrm{e23}}\rangle$ with an equal probability of $\frac{1}{4}$. By comparing with the trivial fusion, the initial ground state can evolve into different excited states or remains at ground states, while it always remains at ground state in the trivial fusion. Therefore, measurements can be carried out through comparative experiments of the trivial and nontrivial fusions. The result can be explained by the transition with different basis vectors~\cite{SM}:
\begin{equation}
	\begin{aligned}
		|0_{12}\tilde{0}_{12}0_{34}\tilde{0}_{34}\rangle&=\frac{1}{2}(|0_{14}\tilde{0}_{14}0_{23}\tilde{0}_{23}\rangle-|1_{14}\tilde{1}_{14}1_{23}\tilde{1}_{23}\rangle)\\
		&-\frac{1}{2}(i|1_{14}\tilde{0}_{14}1_{23}\tilde{0}_{23}\rangle+i|0_{14}\tilde{1}_{14}0_{23}\tilde{1}_{23}\rangle). 
	\end{aligned}
\end{equation}
This can be mapped to a general anyon model by $F^{\sigma\sigma\sigma}_{\sigma}$ and $\tilde{F}^{\tilde{\sigma}\tilde{\sigma}\tilde{\sigma}}_{\tilde{\sigma}}$. Thus, the nontrivial fusion in a time-reversal invariant TSC agrees with the non-Abelian statistics as a progress that the Ising anyons fuse in two different TR sectors independently. Furthermore, the measurement of  $|1_{\mathrm{e23}}\tilde{0}_{\mathrm{e23}}\rangle$ and $|0_{\mathrm{e23}}\tilde{1}_{\mathrm{e23}}\rangle$ can serve as the signal of the emergent supersymmetry.
\begin{figure}
	\centering 
	\includegraphics[width=0.48\textwidth]{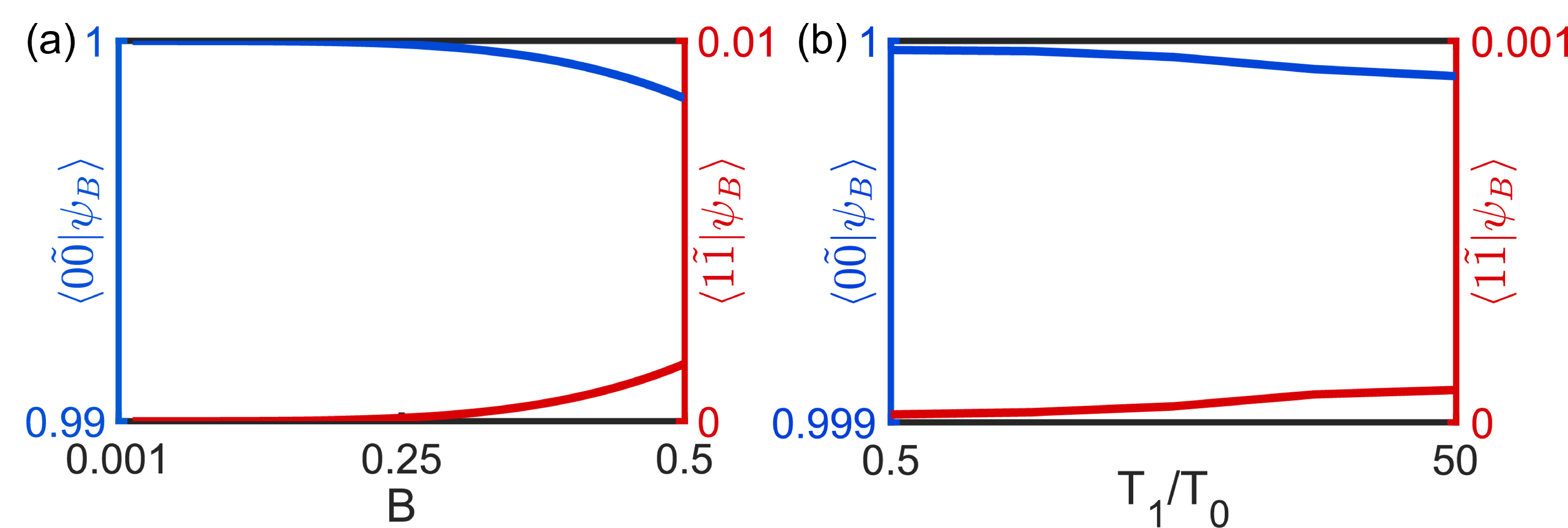}  
	\caption{(a) The mixing as a function of the strength $B$. Increasing $B$ from $0.001$ to $0.5$, a small mixing happens only when $B$ is very large. $t_{1}$ is initially set to 0.001, $T_{0}=200$ and $T_{1}=500$. (b) The mixing as a function of the rate of the mixing. $t_{1}$ is initially set to 0.001 and $B=0.25$.}
	\label{fig3}.
\end{figure}

\textit{Local adiabatic mixing---.} Local adiabatic mixing can mix states with the same fermion-parity as $U|0\tilde{0}\rangle= \cos{\theta}|0\tilde{0}\rangle +\sin{\theta}|1\tilde{1}\rangle$ in an adiabatic progress that always satisfies $\mathcal{T}$~\cite{Wolms2014}. The unitary operator $U=e^{\theta\gamma\tilde{\gamma}}$ is defined to describe the mixing, impacting the braiding of MKPs. This mixing has been shown as a higher-order effect, and some extra unitary symmetries can avoid such mixing~\cite{Xiong-Jun Liu2022,P. Gao2016}. As detailed in the Supplemental Materials~\cite{SM}, the effect of mixing is much smaller in the fusion as a consequence that only has non-negligible effect near the energy-degenerate states, as the dots N and F shown in Figs.~\ref{fig2}(c) and \ref{fig2}(d). To illustrate, we consider the local adiabatic mixing near the points N and F at the initialization. The Hamiltonian describing the initialization and  this effect can be written as $H_{m}=it_{1}(\gamma_{1}\gamma_{2}+\tilde{\gamma}_{1}\tilde{\gamma}_{2})+i1/2\xi(2-\gamma_{3}\gamma_{4}-\tilde{\gamma}_{3}\tilde{\gamma}_{4})+
it_{0}(\gamma_{2}\gamma_{3}+\tilde{\gamma}_{2}\tilde{\gamma}_{3})+it_{00}(\gamma_{2}\tilde\gamma_{3}-\tilde{\gamma}_{2}\gamma_{3})$ with $t_{0}=B\cos\theta$ and $t_{00}=B\sin\theta$, which adds a Kramers pair of Andreev bound states denoted by $f^{\dagger}_{2}=1/2(\gamma_{3}+i\gamma_{4})$, $\tilde{f}^{\dagger}_{2}=1/2(\tilde{\gamma}_{3}+i\tilde{\gamma}_{4})$~\cite{Knapp2020}. In proximity to the energy-degenerate point, assuming the normal coupling $t_{1}$ is evolved to $0$ with the rate $1/T_{0}$ and $\theta$ is evolved from $0$ to $2\pi$ with a rate of $1/T_{1}$, where $T_{0}$ is the time that $t_{1}$ is evolved to $0$ and $T_{1}$ is the period of mixing. The degrees of mixing as functions of the strength $B$ and the rate denoted by $T_{1}/T_{0}$ are displayed in Figs.~\ref{fig3}(a) and \ref{fig3}(b), respectively. A small mixing can be caused when the strength $B$ is greatly larger than the initial normal coupling $t_{1}$ and $T_{0}$ is greatly larger than $T_{1}$. Therefore, the local adiabatic mixing is a higher-order effect and only the high-frequency noise makes sense, compared with the rate of $1/T_{0}$ that can be controlled. This indicates that it is reasonable to neglect the influence of local adiabatic mixing in the fusion.

However, even at the extreme situation where the mixing is considerable, the trivial and nontrivial fusions can also be clearly distinguished. In the trivial case, the final state contains $|0_{23}\tilde{0}_{23}\rangle$ with a probability of $1-a_{0}$ and $|1_{23}\tilde{1}_{23}\rangle$ with a probability of $a_{0}$. And, in the nontrivial case, the final state contains $|0_{23}\tilde{0}_{23}\rangle$ and $|1_{23}\tilde{1}_{23}\rangle$ with probabilities of $\frac{1}{4}\pm a_{1}$, and $|0_{23}\tilde{1}_{23}\rangle$ and $|1_{23}\tilde{0}_{23}\rangle$ with probabilities of $\frac{1}{4}\pm a_{2}$. Here, we assume that $a_{0,1,2}$ describe the mixings in the fusion. Thus, the results of the trivial and nontrivial fusions can be distinguished by the fractional Josephson effect~\cite{Knapp2020, X.-J. Liu2014,F. Zhang2013,F.Zhang2014} or other measurements, since the mixing is at the same fermion-parity state. And, it plays negligible influence in the observation of supersymmetry for mixing the states with same fermion-parity.

\begin{figure}
	\centering 
	\includegraphics[width=0.46\textwidth]{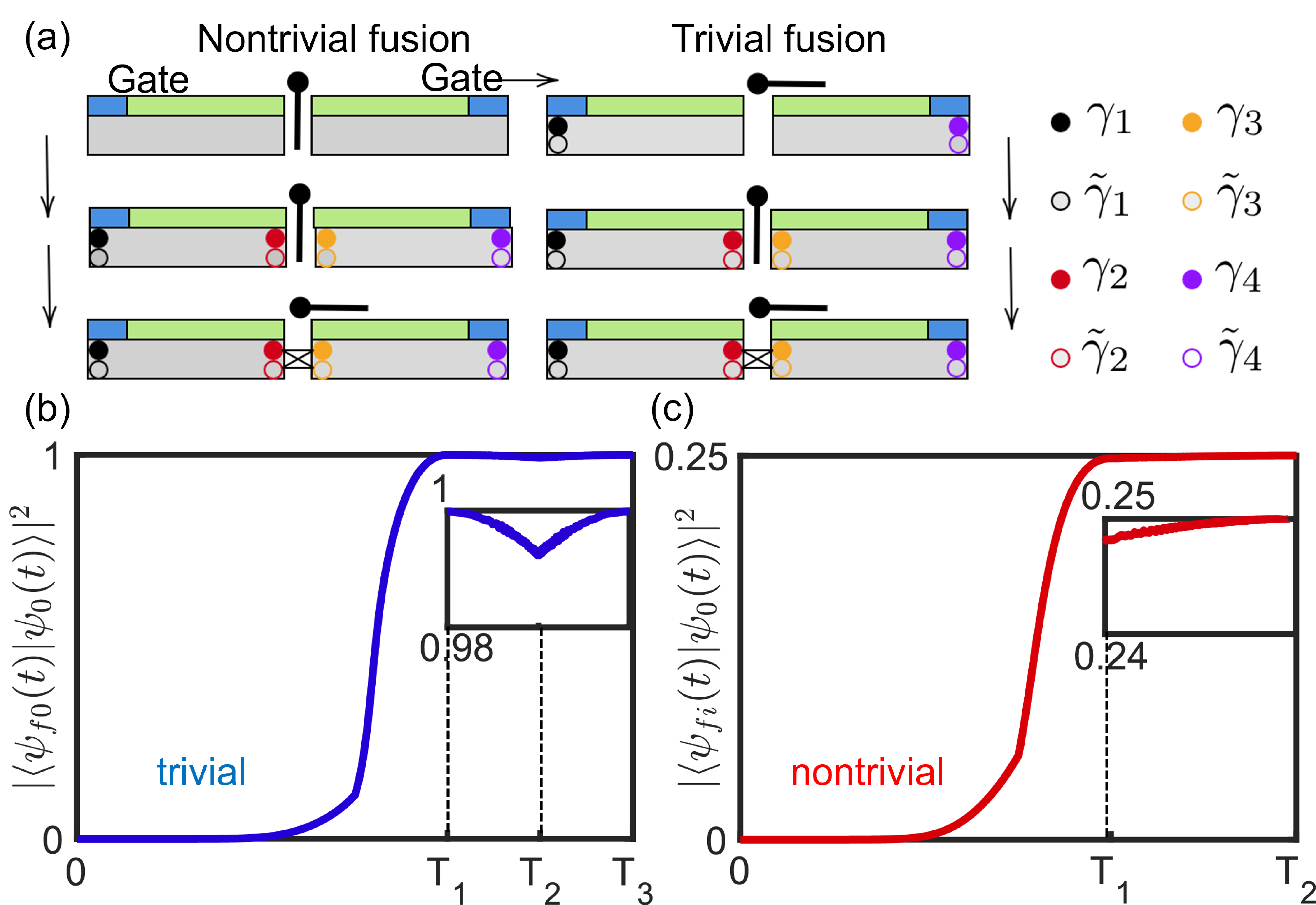} 
	\caption{(a) The scheme of fusion based on $d_{x^{2}-y^{2}}$ paring TSCs. The gray rectangles denote the nanowires and the rectangles with different colors on them denote the gates. (b) The evolution of wave function in the trivial fusion. The initialization is from $0$ to $T_{2}$ and the fusion is from $T_{2}$ to $T_{3}$. (c) The evolution of wave function in the nontrivial fusion. The initialization is from $0$ to $T_{1}$ and the fusion is from $T_{1}$ to $T_{2}$.}
	\label{fig4}.
\end{figure}

\textit{Fusion in $d_{x^2-y^2}$-wave TSCs---.} As a concrete example, a time-reversal-invariant TSC can be produced in a nanowire with $d_{x^2-y^2}$-wave paring and Rashba spin-orbit interaction~\cite{Y. Tanaka2022,C. L. M. Wong2012}. Based on the effective model, two long nanowires connected by a gate can be designed, by using the gates to control the chemical potential $\mu_{1}$ and $\mu_{2}$ at different parts of the two nanowires denoted by blue and green rectangles~\cite{Tong Zhou2022}, respectively, as displayed in Fig.~\ref{fig4}(a). Initially, $\mu_{1}$ is set to be $2$ to make the small regions being topological at the left and right parts of the two nanowires, respectively. Then, the system is at the ground state $\psi_{0}$. In the trivial fusion, using the gate denoted by the green rectangles to make the rest regions of the two nanowires be topological by decreasing the chemical potential $\mu_{2}$ from $8$ to $2$ and connecting the two nanowires by the gate with the coupling from $0$ to $1$. $\gamma_{1,4}$ and $\tilde{\gamma}_{1,4}$ emerge as displayed in Fig.~\ref{fig4}(b). By disconnecting the two TSCs via the gate, the initial state $|0_{23}\tilde{0}_{23}\rangle$ formed by $\gamma_{2,3}$, $\tilde{\gamma}_{2,3}$ can be prepared. By connecting the two nanowires, it returns to the ground state with an absolute probability of $100\%$, shown by $|\langle\psi_{f0}(t)|\psi_{0}(t)\rangle)|^{2}$ in Fig.~~\ref{fig4}(b), where $|\psi_{f0}(t)\rangle$ denotes the ground state at the time $t=T_{3}$. In the nontrivial fusion,  by decreasing $\mu_{2}$ to $2$ and disconnecting the nanowires, $\gamma_{1,2,3,4}$ and $\tilde{\gamma}_{1,2,3,4}$ emerge as displayed in Fig.~\ref{fig4}(a). By further connecting the two nanowires via increasing the coupling to $1$, it returns to $|0_{23}\tilde{0}_{23}\rangle$, $|1_{23}\tilde{1}_{23}\rangle$, $|1_{23}\tilde{0}_{23}\rangle$, and $|0_{23}\tilde{1}_{23}\rangle$ with an equal probability of $25\%$, displayed by $|\langle\psi_{fi}(t)|\psi_{0}(t)\rangle)|^{2}$ in Fig.~\ref{fig4}(c), where $|\psi_{fi}(t)\rangle$ denotes all the instantaneous eigenstates in the gap at $t=T_{2}$.

\textit{Conclusion---.} Braiding and nontrivial fusion are two aspects of the non-Abelian statistics of MKPs. Corresponding to the braiding, MKPs behave as Ising anyons in two time-reversal sectors in the fusion and they fuse into different states with an equal probability in the nontrivial fusion. In general, the local adiabatic mixing is shown to be negligible in the fusion. Furthermore, we show that the measurement of the odd fermion-parity states indicates the sign of supersymmetry. So far, the time-reversal-invariant TSC has yet been observed in experiment. We show that the nontrivial fusion may provide an alternative manner to confirm it. Multiple MZMs can be protected by the unitary symmetry, which can avoid the local adiabatic mixing~\cite{Xiong-Jun Liu2022}. The similar results for the trivial and nontrivial fusion can also be obtained~\cite{SM}, as well as the supersymmetry observation. Thus, our theory provide an ideal platform to study the fusion of multiple MZMs.

\textit{Acknowledgements---.} This work was financially supported by the National Natural Science Foundation of China (Grants No. 11974327, and 12234017), Anhui Initiative in Quantum Information Technologies (AHY170000), and Innovation Program for Quantum Science and Technology (2021ZD0302800). We also thank the Supercomputing Center of University of Science and Technology of China for providing the high performance computing resources.


\begin{thebibliography}{99}
	\bibitem{C. Nayak2008} C. Nayak, S. H. Simon, A. Stern, M. Freedman, and S. Das Sarma, Rev. Mod. Phys. \textbf{80}, 1083 (2008).
	\bibitem{Alicea2011} J. Alicea, Y. Oreg, G. Refael, F. von Oppen, and M. P. A. Fisher, Nat. Phys. \textbf{7}, 412 (2011).
	\bibitem{Alicea2012} J. Alicea, Rep. Prog. Phys. \textbf{75}, 076501 (2012).
	\bibitem{Sato2017} M. Sato and Y. Ando, Rep. Prog. Phys. \textbf{80}, 076501 (2017).
	\bibitem{C.W2013} C. W. J. Beenakker, Annu. Rev. Condens. Matter Phys. \textbf{4}, 113 (2013).
	\bibitem{M.Sato2016} M. Sato and S. Fujimoto, J. Phys. Soc. Jpn. \textbf{85}, 072001
	(2016).
	\bibitem{M. Franz2015} S. R. Elliott and M. Franz, Rev. Mod. Phys. \textbf{87}, 137 (2015).
	\bibitem{X.-J. Liu2020} Y.-P. He, J.-S. Hong, X.-J. Liu, Acta Phys. Sin. \textbf{69(11)}, 110302 (2020).
	\bibitem{Kitaev2001} A. Y. Kitaev, Phys. Usp. \textbf{44}, 131 (2001).
	\bibitem{Green2000} N. Read and D. Green, Phys. Rev. B \textbf{61}, 10267 (2000).
	\bibitem{L. Fu2008} L. Fu, and C. L. Kane, Phys. Rev. Lett. \textbf{100}, 096407 (2008).
	\bibitem{R. M. Lutchyn2010} R. M. Lutchyn, J. D. Sau, and S. Das Sarma, Phys. Rev. Lett. \textbf{105}, 077001 (2010).
	\bibitem{Y. Oreg2010} Y. Oreg, G. Refael, and F. von Oppen, Phys. Rev. Lett. \textbf{105}, 177002 (2010).
	\bibitem{F. Pientka2017} F. Pientka, A. Keselman, E. Berg, A. Yacoby, A. Stern, and B. I.  Halperin, Phys. Rev. X \textbf{7}, 021032 (2017).
	\bibitem{M. Sato2009} M. Sato, Y. Takahashi, and S. Fujimoto, Phys. Rev. Lett. \textbf{103}, 020401 (2009).
	\bibitem{P. W. Brouwer2017} J. Langbehn, Y. Peng, L. Trifunovic, F. von Oppen, and P. W. Brouwer, Phys. Rev. Lett. \textbf{119}, 246401 (2017).
	\bibitem{F. Zhang2018} Q. Wang, C.-C. Liu, Y.-M. Lu, and F. Zhang, Phys. Rev. Lett. \textbf{121}, 186801 (2018).
	\bibitem{C. L. M. Wong2012} C. L. M. Wong and K. T. Law, Phys. Rev. B \textbf{86}, 184516 (2012).
	\bibitem{Qi2009} X.-L. Qi, T. L. Hughes, S. Raghu, and S.-C. Zhang, Phys. Rev. Lett. \textbf{102}, 187001(2009).
	\bibitem{F. Zhang2013} F. Zhang, C. L. Kane, and E. J. Mele, Phys. Rev. Lett. \textbf{111}, 056402 (2013).
	\bibitem{Karsten Flensberg2014} E. Gaidamauskas, J. Paaske, and K. Flensberg, Phys. Rev. Lett. \textbf{112}, 126402 (2014). 
	\bibitem{D. Loss2018} M. Thakurathi, P. Simon, I. Mandal, J. Klinovaja, and D. Loss, Phys. Rev. B \textbf{97}, 045415 (2018).
	\bibitem{E. Berg2013} A. Keselman, L. Fu, A. Stern, and E. Berg, Phys. Rev. Lett. \textbf{111}, 116402 (2013).
	\bibitem{Jelena Klinovaja2020} Y. Volpez, D. Loss, and J. Klinovaja, Phys. Rev. Research \textbf{2}, 023415 (2020).
	\bibitem{Corneliu Malciu2018} C. Malciu, L. Mazza, and C. Mora, Phys. Rev. B \textbf{98}, 165426 (2018).
	\bibitem{Xiaoyu Zhu2018} X.-Y. Zhu, Phys. Rev. B \textbf{97}, 205134 (2018).
	\bibitem{Chetan Nayak2008} P. Bonderson, M. Freedman, and C. Nayak, Phys. Rev. Lett. \textbf{101}, 010501 (2008).
	\bibitem{Liang Fu2022} C. Schrade and L. Fu, Phys. Rev. Lett. \textbf{129}, 227002 (2022).
	\bibitem{J. Liu2021} J. Liu, W. Chen, M. Gong, Y. Wu, and X. C. Xie, Sci. China-Phys. Mech. and Astron. \textbf{64}, 117811 (2021).
	\bibitem{D. Aasen2016} D. Aasen, M. Hell, R. V. Mishmash, A. Higginbotham, J. Danon, M. Leijnse, T. S. Jespersen, J. A. Folk, C. M. Marcus, K. Flensberg, and J. Alicea, Phys. Rev. X \textbf{6}, 031016 (2016).
	\bibitem{X.-J. Liu2014} X.-J. Liu, C. L. M. Wong, and K. T. Law, Phys. Rev. X \textbf{4}, 021018 (2014).
	\bibitem{Y. Tanaka2022} Y. Tanaka, T. Sanno, T. Mizushima, and S. Fujimoto, Phys. Rev. B \textbf{106}, 014522 (2022).
	\bibitem{Tong Zhou2022} T. Zhou, M. C. Dartiailh, M. C. Dartiailh, J. E. Han, A. Matos-Abiague, J. Shabani, I. Žutić, Nat. Commun. \textbf{13}, 1738 (2022).
	\bibitem{E. Altman2015} J. Ruhman, E. Berg, and E. Altman, Phys. Rev. Lett. \textbf{114}, 100401 (2015).
	\bibitem{K. T. Law2009}	K. T. Law, Patrick A. Lee, and T. K. Ng, Phys. Rev. Lett. \textbf{103}, 237001 (2009).
	\bibitem{H.Shtrikman2012} A. Das, Y. Ronen, Y. Most, Y. Oreg, M. Heiblum, and H. Shtrikman, Nat. Phys. \textbf{8}, 887 (2012).
	\bibitem{J. Liu2012}  J. Liu, A. C. Potter, K. T. Law, and P. A. Lee, Phys. Rev. Lett. \textbf{109}, 267002 (2012).
	\bibitem{L. P. Kouwenhoven2012} V. Mourik, K. Zuo, S. M. Frolov, S. R. Plissard, E. P. A. M. Bakkers, and L. P. Kouwenhoven, Science \textbf{336}, 1003 (2012).
	\bibitem{F. von Oppen2107} D. Litinski and F. von Oppen, Phys. Rev. B \textbf{96}, 205413 (2017).
	\bibitem{Wolms2014} K. Wölms, A. Stern, and K. Flensberg, Phys. Rev. Lett. \textbf{113}, 246401 (2014).
	\bibitem{Wolms2016} K. Wölms, A. Stern, and K. Flensberg, Phys. Rev. B \textbf{93}, 045417 (2016).
	\bibitem{P. Gao2016} P. Gao, Y.-P. He, and X.-J. Liu, Phys. Rev. B \textbf{94}, 224509 (2016).
	\bibitem{F.Zhang2014} F. Zhang and C. L. Kane,
	Phys. Rev. B \textbf{90}, 020501(R) (2014).
	\bibitem{B. Béri} E. Mellars and B. Béri, Phys. Rev. B \textbf{94}, 174508 (2016).
	\bibitem{F. von Oppen2017} A. Camjayi, L. Arrachea, A. Aligia, and F. von Oppen,
	Phys. Rev. Lett. \textbf{119}, 046801 (2017).
	\bibitem {Xiong-Jun Liu2022} J.-S. Hong, T.-F. Jeffrey Poon, L. Zhang, and X.-J. Liu, Phys. Rev. B \textbf{105}, 024503 (2022).
	\bibitem {SM} See details in the Supplemental Materials.
	\bibitem{Knapp2020} C. Knapp, A. Chew, and J. Alicea, Phys. Rev. Lett. \textbf{125}, 207002 (2020).
\end{thebibliography}
\end{document}